# Kidney function and kidney failure prediction in a large multiethnic population

James A. Diao, M.D., M.Phil.[a,b,c] *, Morgan Sanchez, B.S.[a,b] *, Adir Sommer, M.D.[d], Keren Cohen, B.S.[d], Maya Makov-Assif, M.D.[d], Marinka Zitnik, Ph.D.[a,b,e,f], Ben Reis, Ph.D.[a,g], Ran D. Balicer, M.D., M.P.H.[a,d,h] **, Noa Dagan, M.D., Ph.D.[a,d,i] **, Arjun K. Manrai, Ph.D.[a,b] **

[a]Ivan and Francesca Berkowitz Family Living Laboratory, Harvard Medical School and Clalit Research Institute, Boston, MA, USA and Tel Aviv, Israel
[b]Department of Biomedical Informatics, Harvard Medical School, Boston, MA, USA
[c]Department of Medicine, Brigham and Women's Hospital, Boston, MA, USA
[d]Clalit Research Institute, Innovation Division, Clalit Health Services, Tel Aviv, Israel
[e]Kempner Institute for the Study of Natural and Artificial Intelligence, Harvard University, Cambridge, MA, USA
[f]Broad Institute of MIT and Harvard, Cambridge, MA, USA
[g]Predictive Medicine Group, Computational Health Informatics Program, Boston Children's Hospital, Boston, MA, USA
[h]School of Public Health, Faculty of Health Sciences, Ben Gurion University of the Negev, Be'er Sheva, Israel
[i]Software and Information Systems Engineering, Ben Gurion University of the Negev, Be'er Sheva, Israel

* Co-first authors
** Co-senior authors

Correspondence:
Arjun K. Manrai, Ph.D.
Department of Biomedical Informatics, Harvard Medical School
10 Shattuck St., Boston, MA, 02115
Arjun_Manrai@hms.harvard.edu

## ABSTRACT

**Background**: Patients with chronic kidney disease (CKD) experience worsening kidney function and develop subsequent kidney failure at different rates. Accurate estimates of a patient's current and future kidney function are needed for optimal clinical decision-making.

**Methods**: To compare current and previously recommended equations for estimating kidney function and risk of kidney failure across a large multiethnic population, we conducted a retrospective multicenter cohort study of primary care, acute care, and hospital settings. Our study population comprised 1,909,042 adults with at least one recorded serum creatinine measurement during 2012-2014, with follow-up until January 2025. Our primary outcomes were the area under the receiver operating characteristic curve and prevalence of CKD by stage across regions of origin.

**Findings**: GFR estimates from the two race-stratified equations—2006 Modification of Diet in Renal Disease (MDRD) and 2009 CKD Epidemiology Collaboration (CKD-EPI)—were similarly calibrated. GFR estimates from the two race-neutral equations diverged, with 2021 European Kidney Function Consortium (EKFC) producing *lower* GFR estimates and 2021 CKD-EPI producing *higher* GFR estimates compared to prior equations. The oldest equation, 2006 MDRD, was the most discriminative of kidney failure within 5 years while the newer European equation, 2021 EKFC, was the least discriminative, with areas under the receiver operating characteristic (AUROC) of 0.862 (95% CI, 0.855-0.869) and 0.846 (95% CI, 0.838-0.853), respectively. The age-adjusted prevalence of CKD varied by the choice of eGFR equation (ranging from 8.5% for 2021 CKD-EPI to 10.8% for EKFC) and across regions of birth (ranging from 8.9% for East Sub-Saharan Africa to 15.3% for South Asia).

**Interpretation**: Using 10 years of follow-up for nearly 2 million individuals, our study demonstrates how newly developed US and European equations diverge from previously recommended equations with broad clinical and epidemiological implications.

## INTRODUCTION

The estimated glomerular filtration rate (eGFR) is a widespread measure of kidney function used to define kidney disease and recommendation thresholds for clinical decision-making. In response to widespread concerns regarding the use of race in estimates of kidney function,[1–5] the US National Kidney Foundation and American Society of Nephrology issued joint recommendations for immediate adoption of a new creatinine-based equation to calculate the eGFR without the use of race.[6,7] The 2021 Chronic Kidney Disease (CKD) Epidemiology Collaboration equation (CKD-EPI),[6] was rapidly and widely adopted in the US.[8] However, European Renal Association leadership has recommended against its implementation, citing several concerns including population differences between the US and Europe.[9] Earlier in 2021, the European Kidney Function Consortium (EKFC) published a separate creatinine-based eGFR equation that is claimed to have improved performance in European populations.[10] The predictive performance and clinical implications of these competing eGFR equations across diverse global populations remains unclear.

With its population diversity across global regions of birth, early digitization of health data, and centralized organization under a universal healthcare system, Israel is a well-suited population for comparative evaluation of clinical equations. In this study, we leverage a longitudinal dataset from the Clalit Health Services (CHS), the largest health management organization in Israel, to evaluate the concordance and predictive performance of two proposed creatinine-based equations: the 2021 CKD-EPI equation and the 2021 EKFC equation. Equations are evaluated in comparison to directly measured creatinine clearance and two previously recommended equations: the 2009 CKD-EPI equation and its predecessor, the 2006 Modification of Diet in Renal Disease (MDRD) equation. Dozens of other equations have been developed and studied,[11] but our study focuses on the ones described above. We studied how the choice of equation affects the relationship between CKD stage and probability of kidney failure, the prediction of kidney function and kidney failure, and the age-adjusted CKD prevalence across regions of birth.

# METHODS

## Setting

CHS serves as both payer and provider of clinical services for the majority of the Israeli population, totaling over 5.4 million patients. We analyzed over 20 years of centralized demographic, diagnostic, laboratory, and administrative data from thousands of clinical sites in CHS. Data from outpatient and inpatient care are collected into a central data warehouse dating back to the 1990s. Since the annual attrition rate of CHS membership is less than 2%, these data spanned more than a decade of consecutive care for nearly all study participants.

## Study population

The main study cohort included CHS participants with at least one serum creatinine measurement within three years prior to January 1, 2015. The collection date of the latest such serum creatinine measurement was defined as the individual's index date, before which covariates are collected and after which follow-up is assessed. As of the index date, participants were required to be aged 25-90 years, have no history of kidney transplant or dialysis, and have continuous CHS membership for at least one year preceding the index date. This main cohort was used to evaluate how the choice of GFR estimating equation may affect CKD prevalence or accuracy in discriminating the 5-year risk of kidney failure. Sub-cohorts were used for additional analyses. A subpopulation of patients with CKD (defined using any of the four evaluated eGFR equations) and with recorded values for the urinary albumin:creatinine ratio (UACR) were used to evaluate associations with incident kidney failure. Another subpopulation of patients with recorded creatinine clearance values was used to cross-tabulate reclassifications between CKD stages when comparing newly developed equations with previously recommended equations and with measured creatinine clearance. We used creatinine clearance as a directly measured surrogate for the GFR, though it is expected to overestimate the GFR owing to tubular secretion of creatinine.[12]

## Variable definitions

Data included demographic information (e.g., age, sex, socioeconomic status, and country of birth), laboratory results (e.g., serum creatinine), diagnoses (e.g., diabetes), and outcomes (e.g., initiation of dialysis, kidney transplantation, or death). All variables and outcomes were accompanied by a date of measurement. Only laboratory measures collected between 0 to 3 years prior to the index date were included in predictive analyses. Serum creatinine values collected during periods of documented or inferred pregnancy were excluded. This study was performed with the approval of the Institutional Review Board at CHS.

## Outcome definitions

We first sought to assess the predictive performance and generalizability of different eGFR equations in the CHS population. Serum creatinine values in the CHS data were derived using the Jaffe method and standardized to isotope dilution mass spectrometry. To avoid an imbalanced comparison where one equation is recalibrated against the Clalit population but its comparator is not, we used the default Q values for EKFC of 0.62 for men and 0.80 for women rather than the observed median serum creatinine values of 0.66 and 0.89. CKD was defined as two or more consecutive measures spanning at least three months of eGFR <60 ml/min/1.73 $m^2$ or UACR ≥30 mg/g, for any time frame prior to the index date.[13] Individuals without data on either urinary albumin or urinary creatinine were assumed to not have albuminuria. We further stratified individuals by CKD stage, ranging from G1 (highest kidney function) to G5 (lowest kidney function). G1 and G2 are defined as eGFR ≥90 ml/min/1.73 $m^2$ and eGFR ≥60 ml/min/1.73 $m^2$, respectively, with UACR ≥30 mg/g. G3a, G3b, G4, and G5 are defined as eGFR less than 60, 45, 30, and 15 ml/min/1.73 $m^2$, respectively. The primary outcome for comparative analyses was 5-year incident kidney failure, defined as either kidney transplantation or initiation of dialysis.

**Statistical analysis**

To study the reclassification of individuals between CKD stages, we cross-tabulated study participants with recorded measured creatinine clearance values across intersections of CKD stages as defined using different eGFR equations. Each individual was assigned to one entry, and the 5-year risk of kidney failure was calculated for each entry based on the Fine-Gray subdistribution hazard function. Comparisons included new and prior US and European equations (2006 MDRD, 2021 CKD-EPI, and 2021 EKFC) with either the widely used 2009 CKD-EPI equation or with measured creatinine clearance. Notably, the 2009 CKD-EPI equation used the same modeling approach as 2006 MDRD and differs only by its inclusion of additional study cohorts, including participants without kidney disease. The 2021 CKD-EPI equation used the same modeling approach and data as 2009 CKD-EPI, but was refit to remove race as an input variable. The European equation (2021 EKFC) differs from the aforementioned US-derived equations in terms of both modeling approach and data.

To investigate the ability of CKD stages to stratify patients according to kidney failure risk, we produced survival curves computing the probability of remaining free from kidney failure over time for each of the four eGFR equations. As before, we calculated the survival function using the Fine-Gray subdistribution hazard function with death modeled as a competing risk.[14] The curves were stratified for different CKD stages, with descent over time representing cumulative events of kidney failure.

We then sought to compare estimates of kidney function by their ability to discriminate patients at higher risk of progressing to kidney failure. First, we computed precision, sensitivity, and specificity values when using a range of eGFR cutoffs to predict kidney failure. We then repeated the same process for a range of cutoffs based on the Kidney Failure Risk Equation (KFRE).[15,16] The four-variable KFRE is an equation validated for the prediction of kidney failure—similarly defined as kidney transplant or initiation of dialysis—using age, sex, eGFR, and UACR among patients with eGFR <60 ml/min/1.73 $m^2$.

Overall discriminative accuracy was measured using the area under the receiver operating characteristic curve (AUROC) and area under the precision-recall curve (AUPRC). An AUROC or AUPRC of 1.0 indicates perfect prediction, while an AUROC of 0.5 or AUPRC equal to the outcome prevalence indicates "null" predictions that are not more accurate than random chance. Because UACR is required for computing the KFRE, this analysis was conducted in the subpopulation of individuals with available UACR measurements.

Lastly, we performed descriptive analyses to characterize the prevalence of CKD as defined using each of the four GFR estimating equations. We stratified our analyses by region of birth, using the Global Burden of Disease (GBD) super regions which were grouped by cause-of-death distributions.[17] We calculated the age-adjusted prevalence of CKD across population subgroups by applying inverse probability weights developed using overall population proportions within age bins, including 25-29 years, 30-34 years, …, 85-90 years.

# RESULTS

**Population description**

The study population comprised 1,908,042 adults with reported creatinine values in the three years spanning 2012-2014 (Table 1). The population was 45.1% male and 54.9% female with a median age of 49.1 years. Continuous uninterrupted follow-up spanning the full study period of 10 years was available for nearly all participants, with a median follow-up duration of 9.8 years. The selected study population was socioeconomically representative of the total CHS population, with 26.9% in the lowest tertile and 37.4% in the highest tertile. The majority of participants (63.5%) were born in Israel, but the remainder included members of a global diaspora. The most common region of birth for immigrants was Central and Eastern Europe and Central Asia (17.0%), followed by North Africa and the Middle East (12.7%). The remaining regions, including East Sub-Saharan Africa, Western Europe, Southern Latin America, Central Asia, High-Income North America, and South Asia, each accounted for over 10,000 participants. Subpopulations defined based on CKD status or recorded creatinine clearance values were associated with increased medical comorbidities and incidence of kidney-related outcomes (Table 1).

**Concordance of CKD stages and kidney failure risk**

We cross-tabulated the intersection of CKD stages defined using alternate eGFR equations as compared to the commonly used 2009 CKD-EPI equation (Figure 1A). We also compared these equations to measured creatinine clearance in terms of the risk of kidney failure at each cross-tabulated CKD stage (Figure 1B). Entries along the diagonal represent concordance, while entries farther off the diagonal represent increasing discordance. Relative to the previously recommended 2009 CKD-EPI equation, using 2021 CKD-EPI results in net reclassification to less severe CKD stages (higher eGFR). By contrast, using 2021 EKFC results in net reclassification to more severe CKD stages (lower eGFR) and using 2006 MDRD results in similar proportions reclassified to more or less severe CKD stages, indicating similar calibration of eGFR values.

The probability of kidney failure increases with CKD stage as defined using any of the eGFR equations or using measured creatinine clearance (Figure 1B). When individuals are assigned discordant CKD stages by an eGFR equation and creatinine clearance, their kidney failure risk was more similar to that of individuals concordantly assigned to the CKD stage assigned by measured creatinine clearance rather than to individuals concordantly assigned to the CKD stage assigned using eGFR based on serum creatinine. This is consistent with an understanding of creatinine clearance as a more direct measurement of underlying kidney function than serum-based estimates such as the eGFR.

**Associations with longitudinal risk of kidney failure**

Figure 2 presents survival curves for kidney failure across CKD stage and choice of GFR estimating equation, as calculated using the Fine-Gray subdistribution hazard function and accounting for death as a competing risk. For each equation, CKD stage successfully stratifies between low-risk and high-risk patients. When inspecting more granular differences between the equations, we note that the average risk of kidney failure among participants in the same CKD stage was typically similar between 2009 CKD-EPI and 2021 CKD-EPI, and divergent for 2006 MDRD and 2021 EKFC (higher and lower rates of kidney failure respectively).

**Prediction of kidney failure risk by eGFR equation and by KFRE**

When evaluating the predictive performance of the eGFR equations, several trends were noted (Figure 3). The oldest equation, 2006 MDRD, was the highest performing equation as assessed by both AUROC and AUPRC (Figure 3). By contrast, the newer EKFC equation was found to have the lowest discriminative accuracy. These trends were observed both when eGFR was used as a single predictor and when it was integrated into the KFRE. The non-GFR features used in the KFRE, including age, sex, and UACR, meaningfully improved predictive performance to AUROC values to approximately 90%, with an average AUROC increase of 4.4 percentage points compared to prediction with eGFR alone.

**Association of CKD prevalence with region of birth**

The age-adjusted prevalence of kidney disease in the adult Clalit population was comparable to the global all-age crude prevalence of 9.1%[13] but varied substantially by the choice of eGFR equation and by region of birth (Figure 4). On average, age-adjusted CKD prevalence was 8.5% with 2021 CKD-EPI, 9.4% with 2009 CKD-EPI, 10.1% with 2006 MDRD, and 10.8% with 2021 EKFC. The relative prevalence differences between eGFR equations was preserved when prevalence was stratified by area of birth, but these prevalence values varied substantially across regions. For example, participants born in East Sub-Saharan Africa had the lowest average age-adjusted prevalence of CKD (8.9%) while participants born in South Asia had the highest age-adjusted prevalence (15.3%). The differences in CKD prevalence produced between estimating equations also varied by region of birth, ranging from 0.69 percentage points in East Sub-Saharan Africa to 5.8 percentage points in Central Europe.

## DISCUSSION

The development and adoption of race-neutral equations for estimating kidney function marks the most significant shift in GFR estimation over the past decade. Our study leveraged harmonized data from nearly two million adults to investigate the application and generalizability of new kidney function equations across a diverse multiethnic population. We found that the choice of eGFR equation has little effect on the prediction of incident kidney outcomes, but has significant implications for the correspondence between CKD stage and the likelihood of kidney failure and the prevalence of CKD across populations. These implications arise from differences in calibration: eGFR calculated using the two race-stratified equations (2006 MDRD and 2009 CKD-EPI) were similarly calibrated, while eGFR calculated using the two race-neutral equations diverged—2021 EKFC yielded lower GFR estimates and 2021 CKD-EPI yielded higher GFR estimates.

Among the four creatinine-based eGFR equations, the oldest equation, the 2006 MDRD equation, demonstrated the highest accuracy in discriminating incident kidney failure. This finding is concordant with several single-site studies,[18,19] but discordant with a larger multinational study showing that 2009 CKD-EPI produced a positive net reclassification improvement (NRI) compared to 2006 MDRD with respect to end-stage kidney disease.[20] This may be explained by the choice of outcome variable—AUROC involves continuous comparison across the range of possible values, while NRI involves discrete comparison across predefined classes; NRI has also been shown to produce false positives with less precise prediction models.[21,22] Further research is needed to identify additional population or analytical differences that may explain this discrepancy. Our comparisons of 2009 CKD-EPI and 2021 CKD-EPI were concordant with prior studies, including analyses of 1.6 million Swedish adults with and without CKD[23] and of 3,873 American adults with mild-to-moderate CKD.[24] These studies also found no difference in discrimination for kidney failure between 2009 CKD-EPI and 2021 CKD-EPI, either alone or as an input for KFRE, but neither included 2006 MDRD or 2021 EKFC for comparison.

Iterative refinements to creatinine-based GFR estimation equations over the past several decades have included individuals with higher glomerular filtration rates in development data and improved accuracy as compared to measured GFR.[25] However, our study suggests that these changes have yielded only marginal improvements in prognostic capability when used alone (AUC differences ≤0.016) or in conjunction with other common predictors (AUC differences ≤0.004). It is possible that this plateau represents the limits of prediction attainable using serum creatinine, highlighting the need for further study of alternative biomarkers, either alone or in conjunction with serum creatinine.

The superior performance of the KFRE illustrates another approach: integration of quantitative albuminuria alongside traditional parameters (age, sex, serum creatinine), with direct optimization for kidney failure prediction, yielded a several-fold greater improvement (mean of 0.044) compared to the maximum discriminative accuracy difference across eGFR equations based on serum creatinine (0.016). As our understanding of the physiology and measurement of novel filtration markers continues to mature, we expect that there will be further opportunities to incorporate additional markers into a more comprehensive understanding of kidney function.

Our findings also reveal substantial differences in the age-adjusted prevalence of CKD based on the choice of eGFR equation, which rivaled the prevalence differences observed across regions of birth. This suggests that the choice of equation is likely to have differential epidemiological impacts across population subgroups. For example, findings of decreased CKD prevalence and smaller differences across eGFR equations among individuals born in East Sub-Saharan Africa, and opposite findings among those born in South Asia and Central

Europe, warrant further investigation into potential environmental, genetic, and socioeconomic contributors. Beyond CKD prevalence, these differences have several implications for clinical decision-making associated with GFR thresholds, including referral to specialist care, listing for kidney transplant, or dosing of renally cleared medications.[26] Such differences therefore prompt further evaluation of differential impacts across population subgroups when new equations are considered for recommendation or implementation.

This study has limitations. First, defining kidney failure outcomes using renal replacement therapy may overlook other clinically meaningful changes, such as large decreases in GFR or cases where treatment is indicated but not offered or accepted. Second, analyses performed on subpopulations defined by the availability of laboratory measurements may be subject to selection bias. Third, creatinine clearance is a widely used but imperfect surrogate for measured GFR,[12] and we lacked data to assess other measures of kidney function, including those based on exogenous filtration markers.

Our findings indicate that the current choice faced by health systems between available eGFR equations has substantial clinical and epidemiological implications. Regardless of the choice of equation, rigorous data on the associations between eGFR and long-term outcomes, such as those presented in this study, will be important for optimizing the care of patients with kidney disease.

## REFERENCES

1 Eneanya ND, Yang W, Reese PP. Reconsidering the Consequences of Using Race to Estimate Kidney Function. *JAMA* 2019; **322**: 113–4.

2 Vyas DA, Eisenstein LG, Jones DS. Hidden in Plain Sight — Reconsidering the Use of Race Correction in Clinical Algorithms. *N Engl J Med* 2020; published online June 17. DOI:10.1056/NEJMms2004740.

3 Diao JA, Inker LA, Levey AS, Tighiouart H, Powe NR, Manrai AK. In search of a better equation - performance and equity in estimates of kidney function. *N Engl J Med* 2021; **384**: 396–9.

4 Powe NR. Black Kidney Function Matters: Use or Misuse of Race? *JAMA* 2020; **324**: 737–8.

5 Ahmed S, Nutt CT, Eneanya ND, *et al.* Examining the Potential Impact of Race Multiplier Utilization in Estimated Glomerular Filtration Rate Calculation on African-American Care Outcomes. *J Gen Intern Med* 2020; published online Oct 15. DOI:10.1007/s11606-020-06280-5.

6 Inker LA, Eneanya ND, Coresh J, *et al.* New Creatinine- and Cystatin C-Based Equations to Estimate GFR without Race. *N Engl J Med* 2021; **385**: 1737–49.

7 Delgado C, Baweja M, Crews DC, *et al.* A Unifying Approach for GFR Estimation: Recommendations of the NKF-ASN Task Force on Reassessing the Inclusion of Race in Diagnosing Kidney Disease. *J Am Soc Nephrol* 2021; **32**: 2994.

8 Genzen JR, Souers RJ, Pearson LN, *et al.* Reported Awareness and Adoption of 2021 Estimated Glomerular Filtration Rate Equations Among US Clinical Laboratories, March 2022. *JAMA* 2022; **328**: 2060–2.

9 Gansevoort RT, Anders H-J, Cozzolino M, *et al.* What should European nephrology do with the new CKD-EPI equation? *Nephrol Dial Transplant* 2023; **38**: 1–6.

10 Pottel H, Björk J, Courbebaisse M, *et al.* Development and Validation of a Modified Full Age Spectrum Creatinine-Based Equation to Estimate Glomerular Filtration Rate : A Cross-sectional Analysis of Pooled Data. *Ann Intern Med* 2021; **174**: 183–91.

11 Earley A, Miskulin D, Lamb EJ, Levey AS, Uhlig K. Estimating equations for glomerular filtration rate in the era of creatinine standardization: a systematic review. *Ann Intern Med* 2012; **156**: 785–95.

12 Shemesh O, Golbetz H, Kriss JP, Myers BD. Limitations of creatinine as a filtration marker in glomerulopathic patients. *Kidney Int* 1985; **28**: 830–8.

13 Stevens PE, Ahmed SB, Carrero JJ, *et al.* KDIGO 2024 Clinical Practice Guideline for the Evaluation and Management of Chronic Kidney Disease. *Kidney Int* 2024; **105**: S117–314.

14 Fine JP, Gray RJ. A proportional hazards model for the subdistribution of a competing risk. *J Am Stat Assoc* 1999; **94**: 496.

15 Tangri N, Stevens LA, Griffith J, *et al.* A predictive model for progression of chronic kidney disease to kidney failure. *JAMA* 2011; **305**: 1553–9.

16 Tangri N, Grams ME, Levey AS, *et al.* Multinational Assessment of Accuracy of Equations for Predicting Risk of Kidney Failure: A Meta-analysis. *JAMA* 2016; **315**: 164–74.

17 GBD 2021 Causes of Death Collaborators. Global burden of 288 causes of death and life expectancy decomposition in 204 countries and territories and 811 subnational locations, 1990-2021: a systematic analysis for the Global Burden of Disease Study 2021. *Lancet* 2024; published online April 3. DOI:10.1016/S0140-6736(24)00367-2.

18 Nyman U, Grubb A, Sterner G, Björk J. The CKD-EPI and MDRD equations to estimate GFR. Validation in the Swedish Lund-Malmö Study cohort. *Scand J Clin Lab Invest* 2011; **71**: 129–38.

19 Chen L-I, Guh J-Y, Wu K-D, *et al.* Modification of diet in renal disease (MDRD) study and CKD epidemiology collaboration (CKD-EPI) equations for Taiwanese adults. *PLoS One* 2014; **9**: e99645.

20 Matsushita K, Mahmoodi BK, Woodward M, *et al.* Comparison of risk prediction using the CKD-EPI equation and the MDRD study equation for estimated glomerular filtration rate. *JAMA* 2012; **307**: 1941–51.

21 Hilden J, Gerds TA. A note on the evaluation of novel biomarkers: do not rely on integrated discrimination improvement and net reclassification index. *Stat Med* 2014; **33**: 3405–14.

22 Pepe MS, Fan J, Feng Z, Gerds T, Hilden J. The Net Reclassification Index (NRI): A misleading measure of prediction improvement even with independent test data sets. *Stat Biosci* 2015; **7**: 282–95.

23 Fu EL, Coresh J, Grams ME, *et al.* Removing race from the CKD-EPI equation and its impact on prognosis in a predominantly White European population. *Nephrol Dial Transplant* 2023; **38**: 119–28.

24 Bundy JD, Mills KT, Anderson AH, *et al.* Prediction of end-stage kidney disease using estimated glomerular filtration rate with and without race : A prospective cohort study: A prospective cohort study. *Ann Intern Med* 2022; **175**: 305–13.

25 Levey AS, Stevens LA, Schmid CH, *et al.* A new equation to estimate glomerular filtration rate. *Ann Intern Med* 2009; **150**: 604–12.

26 Diao JA, Wu GJ, Taylor HA, *et al.* Clinical Implications of Removing Race From Estimates of Kidney Function. *JAMA* 2021; **325**: 184–6.

**Table 1. Demographic, clinical, and laboratory characteristics in the study population**

| Variable | Total study population | Population with CKD | CKD and Recorded UACR | Recorded creatinine clearance |
|---|---|---|---|---|
| **Data** | | | | |
| Count–no. | 1,909,042 | 284,399 | 191,370 | 18,202 |
| Median Follow-up–years | 9.8 | 9.5 | 9.5 | 9.5 |
| **Age, years** | | | | |
| Aged 25-44–% | 37.5 | 3.5 | 4.3 | 13.4 |
| Aged 45-64–% | 35.1 | 21.8 | 25.4 | 27.5 |
| Aged 65-74–% | 14.1 | 24.5 | 25.7 | 26.5 |
| Aged 75 or older–% | 13.3 | 50.1 | 44.6 | 32.6 |
| **Sex** | | | | |
| Male–% | 45.1 | 45.9 | 49.5 | 54.6 |
| Female–% | 54.9 | 54.1 | 50.5 | 45.4 |
| **Region of Birth** | | | | |
| Israel–% | 63.5 | 36.3 | 39.0 | 48.6 |
| Central/Eastern Europe and Central Asia–% | 17.0 | 30.3 | 27.3 | 25.3 |
| North Africa and Middle East–% | 12.7 | 25.7 | 26.4 | 19.2 |
| East Sub-Saharan Africa–% | 2.3 | 1.1 | 1.4 | 0.8 |
| Other–% | 4.5 | 6.5 | 5.9 | 6.1 |
| **Socioeconomic Status** | | | | |
| Low–% | 26.9 | 27.5 | 30.7 | 20.1 |
| Middle–% | 35.7 | 36.6 | 37.6 | 36.6 |
| High–% | 37.4 | 35.9 | 31.7 | 43.3 |
| **Disease Conditions** | | | | |
| Diabetes–% | 19.0 | 51.7 | 65.9 | 46.8 |
| Hypertension–% | 34.1 | 81.3 | 84.4 | 75.4 |
| Chronic Kidney Disease (any equation)–% | 15.1 | 100.0 | 100.0 | 63.8 |
| **Renal Measurements** | | | | |
| Median Serum Creatinine (IQR)–mg/dL | 0.79 (0.26) | 1.03 (0.41) | 0.98 (0.47) | 1.11 (0.63) |
| Recorded CrCl–% | 0.97 | 4.1 | 4.5 | 100.0 |
| Median CrCl (IQR)–ml/min/1.73 $m^2$ | 71.3 (60.0) | 55.6 (43.7) | 55.7 (45.7) | 71.8 (59.9) |
| Recorded UACR–% | 27.6 | 70.5 | 100.0 | 61.5 |
| UACR >30 mg/g–% | 7.5 | 49.7 | 73.8 | 33.5 |
| **Renal Outcomes During Follow-up** | | | | |
| Initiated Dialysis–% | 0.77 | 3.97 | 4.81 | 8.9 |
| Kidney Transplant–% | 0.03 | 0.13 | 0.13 | 0.51 |
| Death–% | 13.9 | 47.5 | 46.5 | 40.7 |

Summary statistics for 1.9 million adults with at least one serum creatinine measurement between 2012-2014, and with age 25-90 years, no pregnancy, and no history of dialysis or kidney transplantation at the index date, among a total population of 5.4 million. Hypertension was defined as a blood pressure above 130/80 on two consecutive occasions prior to the index date, or filling of a prescription for an antihypertensive medication within 3 years prior to the index date. Chronic kidney disease is defined as two or more consecutive measurements of eGFR < 60 ml/min/1.73 $m^2$ (when calculated using any of the four equations: 2021 CKD-EPI, 2009 CKD-EPI, 2006 MDRD, or 2021 EKFC) or UACR > 30 mg/g, spanning at least three months for any interval prior to the index date. Abbreviations: CKD: chronic kidney disease, CKD-EPI: Chronic Kidney Disease Epidemiology Collaboration, CrCl: creatinine clearance, eGFR: estimated glomerular filtration rate, EKFC: European Kidney Function Consortium, IQR: Interquartile range, MDRD: Modification of Diet in Renal Disease, UACR: urinary albumin:creatinine ratio.

**Figure 1. Discordance in CKD Stage Classification and Kidney Failure Risk**

## A. Reclassification from 2009 CKD-EPI to alternate eGFR equations

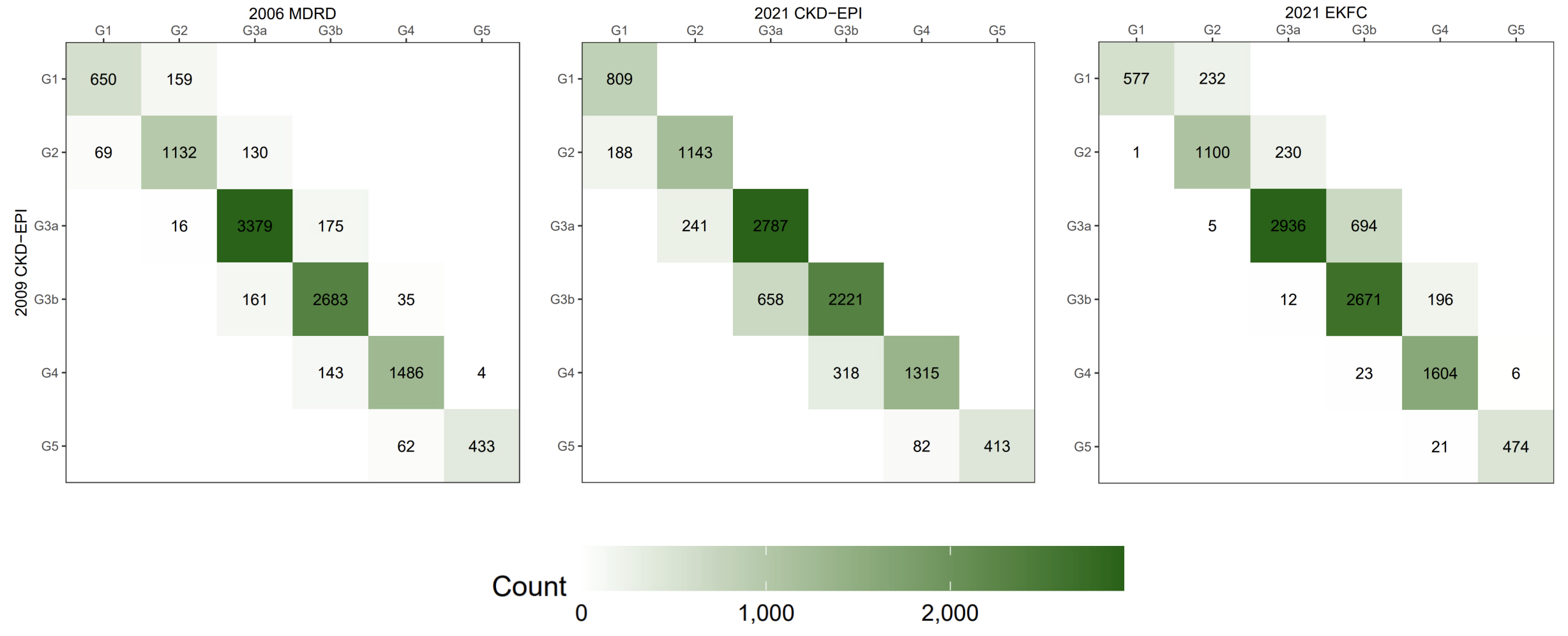

## B. Kidney failure risk of alternate eGFR equations against creatinine clearance

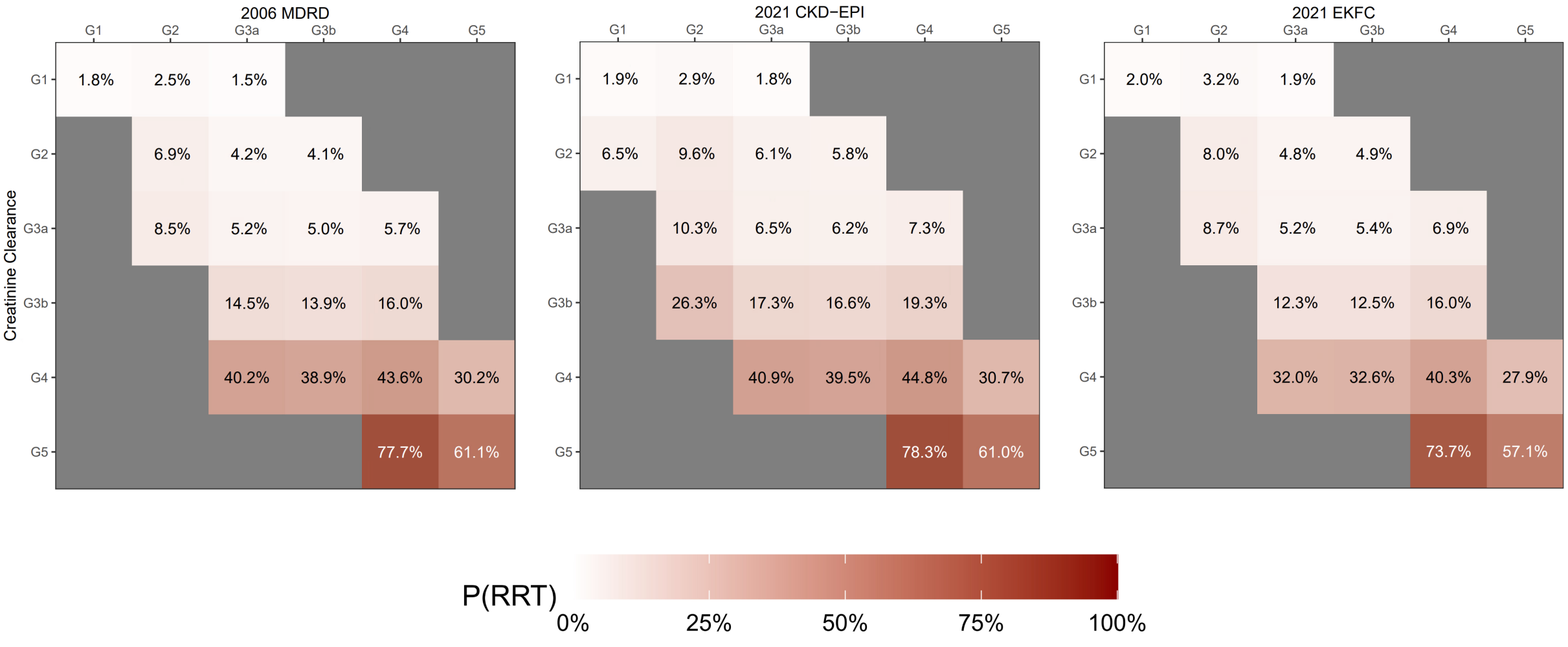

**A.** Classification of chronic kidney disease (CKD) stages when GFR is estimated using the 2009 Chronic Kidney Disease Epidemiology Collaboration (CKD-EPI) or alternate eGFR equations, colored and labeled by the 5-year risk of kidney failure requiring incident dialysis or kidney transplantation. Alternate equations included the earlier 2006 Modification of Diet in Renal Disease equation (2006 MDRD), the newly developed US equation (2021 CKD-EPI), and the newly developed European equation (2021 European Kidney Function Consortium, EKFC). The population comprised 18,202 participants with measured creatinine clearance values. Cells along the top-left to bottom-right diagonal represent concordance, while other cells represent discordance. The GFR thresholds separating G1, G2, G3a, G3b, G4, and G5 are 90, 60, 45, 30, and 15 ml/min/1.73 $m^2$; G1 and G2 additionally require evidence of albuminuria.
**B.** Same as A, but showing risk of kidney failure calculated from the Fine-Gray subdistribution hazard function to account for the competing risk of death. Cross-tabulations are all performed against measured creatinine clearance and each participant is assigned to one cell.

**Figure 2. Rates of Kidney Failure by CKD Stage and GFR Estimating Equation among Participants with Chronic Kidney Disease**

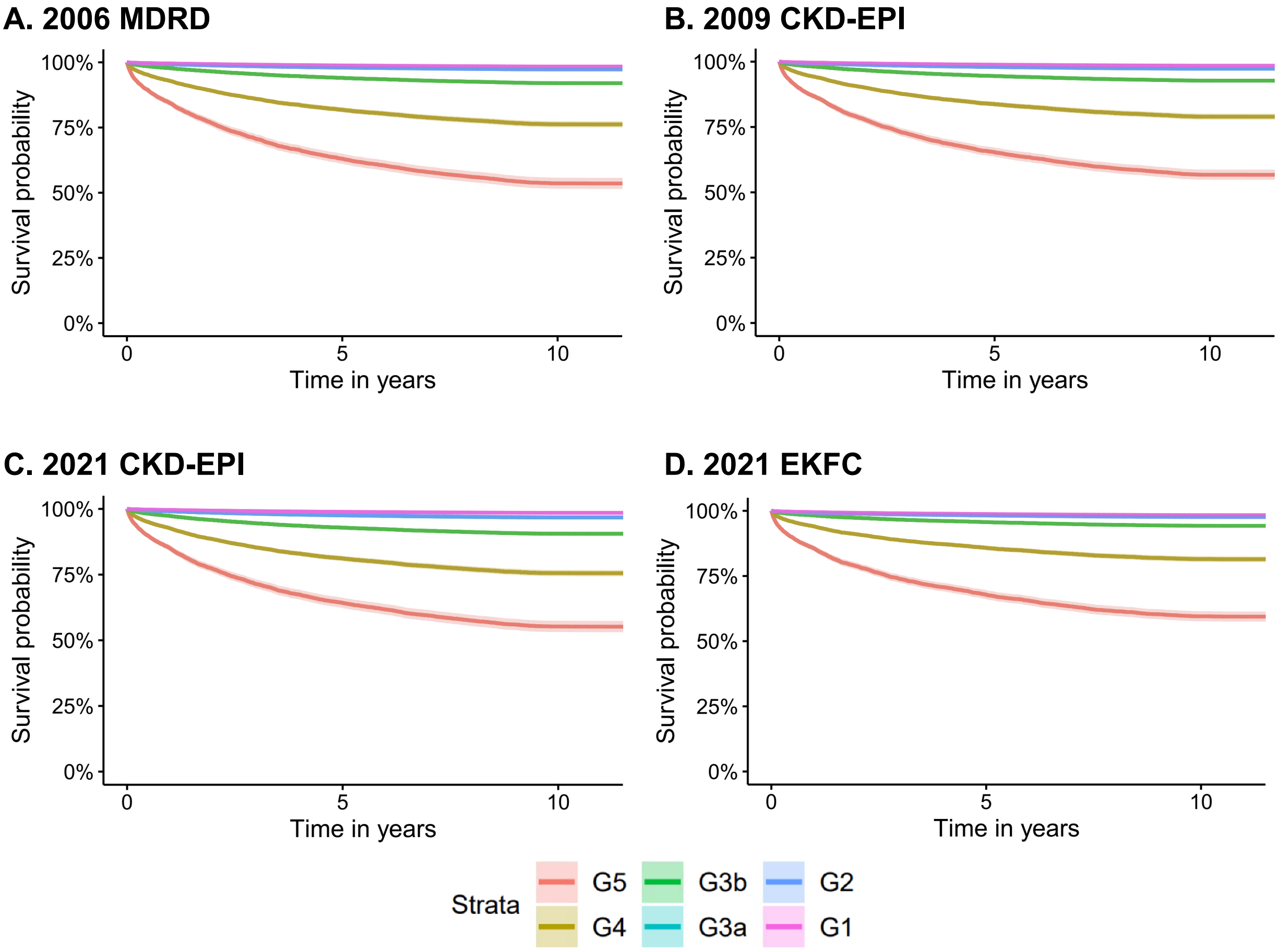

**A.** Survival curves for each of six chronic kidney disease (CKD) stages as calculated using the 2006 Modification of Diet in Renal Disease (MDRD) equation among 284,399 participants with chronic kidney disease. The survival function was calculated using the Fine-Gray subdistribution hazard function to account for the competing risk of death.
**B.** Same as A, but for the 2009 Chronic Kidney Disease Epidemiology Collaboration (CKD-EPI) equation.
**C.** Same as A, but for the 2021 CKD-EPI equation.
**D.** Same as A, but for the 2021 European Kidney Function Consortium (EKFC) equation.

**Figure 3. Prediction of Kidney Failure Risk by eGFR Equation among Patients with CKD and Measured Urinary Albumin:Creatinine Ratio**

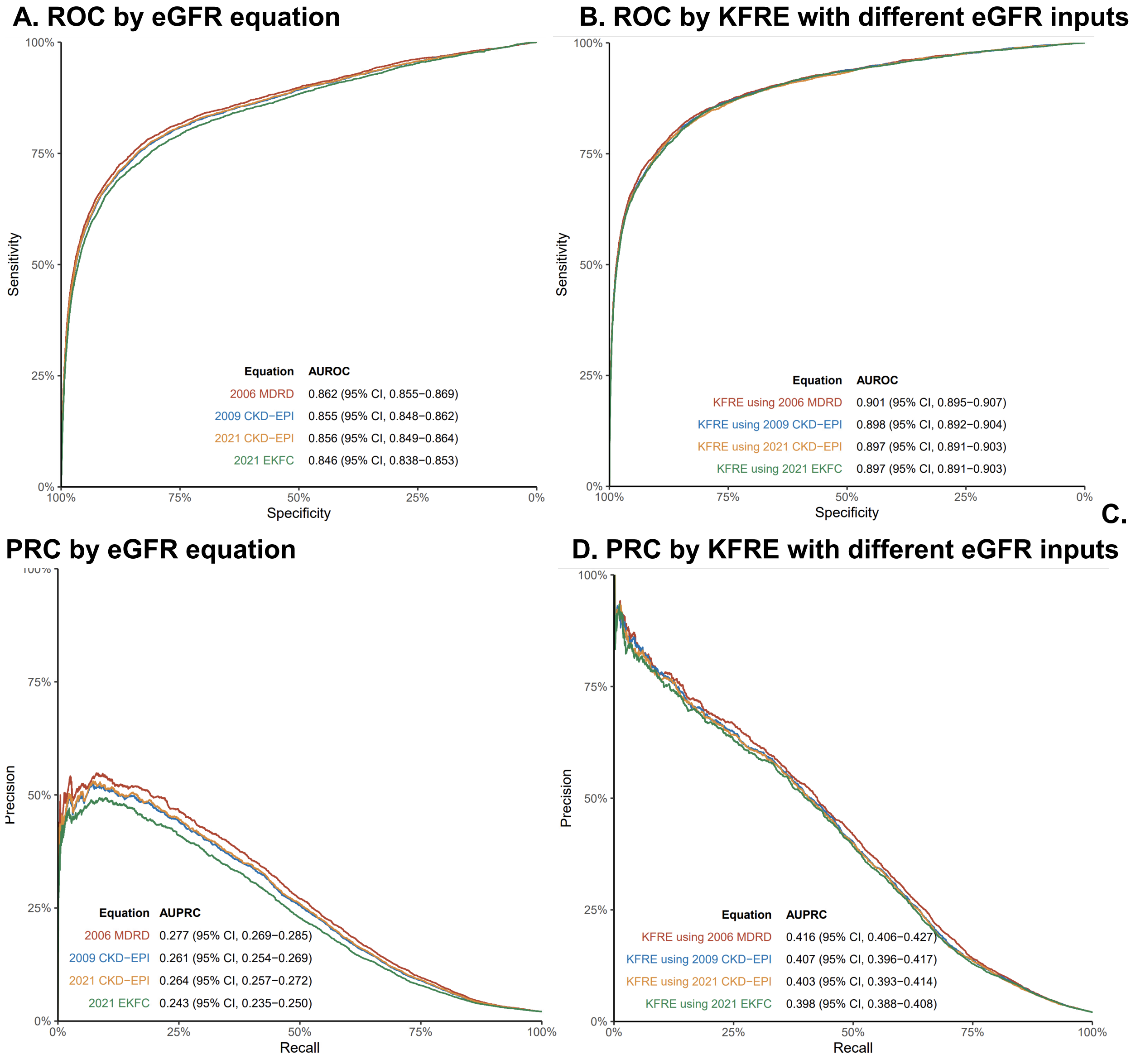

**A.** Receiver operator characteristic (ROC) curves showing sensitivity and specificity values when using four equations for estimated glomerular filtration rates (eGFR) to predict 5-year risk of kidney failure from the date of serum creatinine measurement. Population includes 182,034 adults with at least 5 years of follow-up data without disenrollment or death and data on age, sex, serum creatinine, and urinary albumin:creatinine ratio (UACR), and chronic kidney disease (CKD) as determined by urine albumin:creatinine ratio (UACR) ≥30 mg/g or eGFR <60 ml/min/1.73 $m^2$ as calculated using any of the four eGFR equations. AUROC: area under the ROC curve, CKD-EPI: Chronic Kidney Disease Epidemiology Collaboration, EKFC: European Kidney Function Consortium, MDRD: Modification of Diet in Renal Disease.
**B.** Same as A, but using the Kidney Failure Risk Equation (KFRE) with each eGFR equation.
**C.** Same as A, but for precision-recall curves (PRC). The null model is represented by a horizontal line at the outcome prevalence.
**D.** Same as A, but using the KFRE with each eGFR equation.

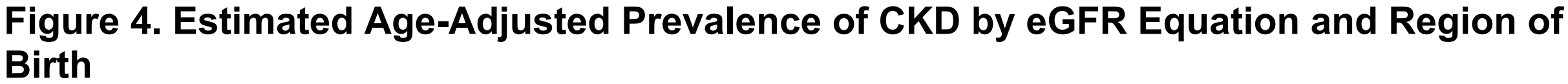

**Figure 4. Estimated Age-Adjusted Prevalence of CKD by eGFR Equation and Region of Birth**

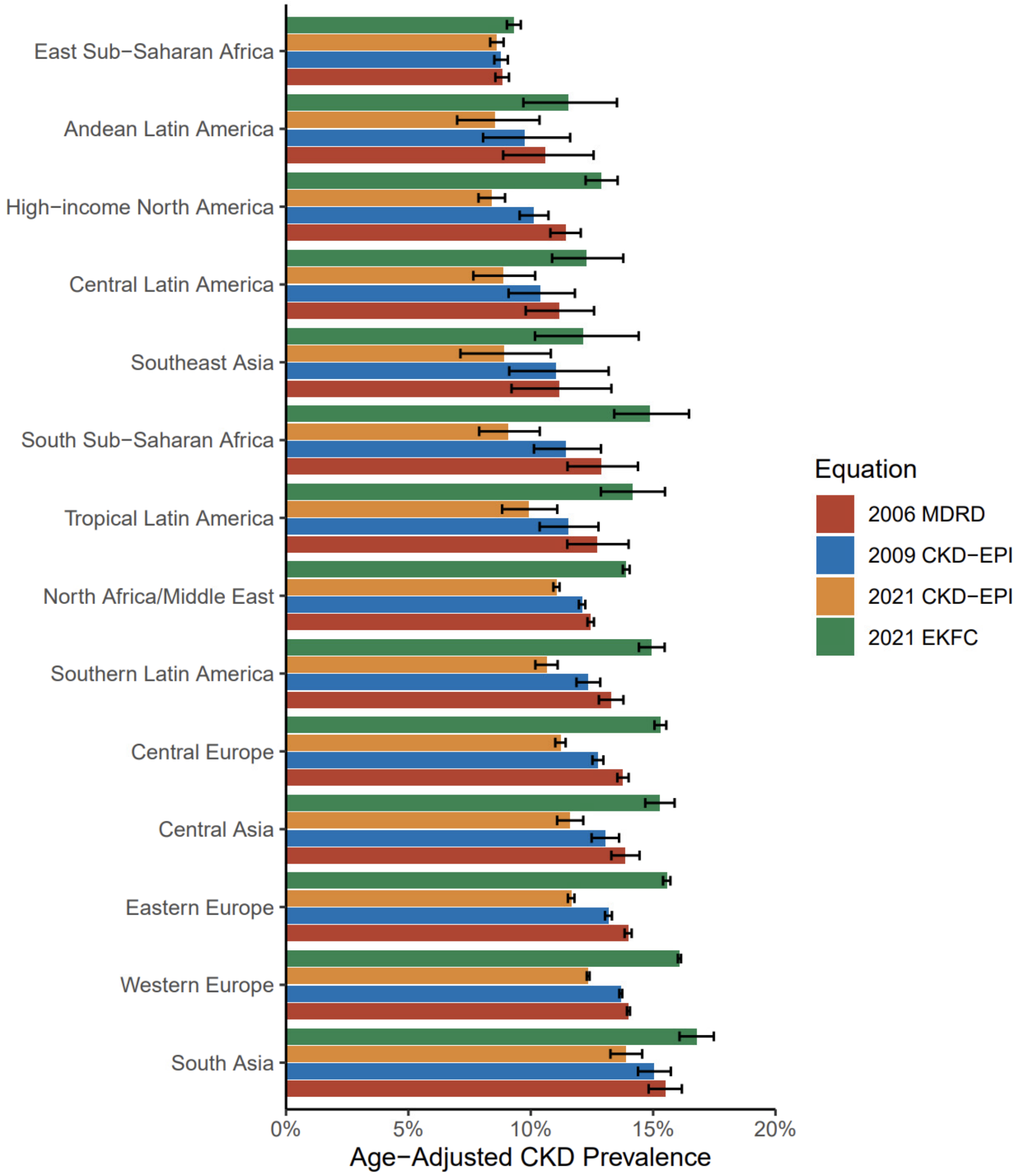

Age-adjusted prevalence of chronic kidney disease (CKD) by Global Burden of Disease (GBD) super region of birth. Chronic kidney disease is defined as two or more consecutive measurements of eGFR < 60 ml/min/1.73 $m^2$ spanning at least three months or most recent urine albumin:creatinine ratio ≥30 mg/g, with both criteria evaluated at any time point prior to the index date. Age adjustment was performed using direct standardization by comparing 5-year age groups (e.g., 25-29 years, 30-34 years, …, 85-90 years) between the population of each country of birth and the total population to account for differences in age distribution. We used “non-Black” race as the input for all race-stratified equations. Regions of birth were ordered by CKD prevalence as defined using the 2009 CKD-EPI equation.